# Multi-mem behavior at reduced voltages in La$_{1/2}$Sr$_{1/2}$Mn$_{1/2}$Co$_{1/2}$O$_{3-x}$ perovskite modified with Sm:CeO$_2$

Wilson Román Acevedo[1], Myriam H. Aguirre[2,3,4], Beatriz Noheda[5,6], Diego Rubi[1,*]

[1] Instituto de Nanociencia y Nanotecnología (INN), CONICET-CNEA, Gral. Paz 1499, 1650 San Martín, Argentina

[2] Instituto de Nanociencia y Materiales de Aragón (INMA-CSIC), Campus Rio Ebro C/Mariano Esquillor s/n, 50018 Zaragoza, Spain

[3] Dpto. de Física de la Materia Condensada, Universidad de Zaragoza, Pedro Cerbuna 12, 50009 Zaragoza, Spain

[4] Laboratorio de Microscopías Avanzadas, Edificio I+D, Campus Rio Ebro C/Mariano Esquillor s/n, 50018 Zaragoza, Spain

[5] CogniGron - Groningen Cognitive Systems and Materials Center, University of Groningen (RuG), Nijenborgh 4, 9747AG Groningen, The Netherlands

[6] Zernike Institute for Advanced Materials, University of Groningen, Nijenborgh 4, 9747AG Groningen, The Netherlands

*Corresponding author: diego.rubi@gmail.com

The use of Machine Learning algorithms is exponentially growing and concerns are being raised about their sustainability. Neuromorphic computing aims to mimic the architecture and the information processing mechanisms of the mammalian brain, appearing as the only avenue that offers significant energy savings compared to the standard digital computers. Memcapacitive devices –which can change their capacitance between different non-volatile states upon the application of electrical stimulation- can significantly reduce the energy consumption of bio-inspired circuitry. In the present work, we study the multi-mem (memristive and memcapacitive) behavior of devices based on thin films of the topotactic redox $La_{1/2}Sr_{1/2}Mn_{1/2}Co_{1/2}O_{3-x}$ (LSMCO) perovskite modified with $Sm:CeO_2$ (SCO), grown on $Nb:SrTiO_3$ with (001) and (110) out-of-plane orientations. Either the self-assembling at the nanoscale of both LSMCO and SCO phases or the doping with Ce(Sm) of the LSMCO perovskite were observed for different fabrication conditions and out-of-plane orientations. The impact of these changes on the device electrical behavior was determined. The optimum devices resulted those with (110) orientation and Ce(Sm) doping the perovskite. These devices displayed a multi-mem behavior with robust memcapacitance and significantly lower operation voltages (especially the RESET voltage) in comparison with devices based on pristine LSMCO. In addition, they were able to endure electrical cycling -and the concomitant perovskite topotactic redox transition between oxidized and reduced phases- without suffering nanostructural or chemical changes. We link these properties to an enhanced perovskite reducibility upon Ce(Sm) doping. Our work contributes to increase the reliability of LSMCO-based multi-mem systems and to reduce their operating voltages closer to the 1 V threshold, which are key issues for the development of nanodevices for neuromorphic or in-memory computing.

I. INTRODUCTION

The development and use of Artificial Intelligence (AI) and Machine Learning (ML) algorithms is growing at vertiginous rates, as they are applied nowadays to solve a large variety of scientific and technological problems [1]. ML algorithms, including Neural Networks (NN), usually run in standard digital computers, where the Central Processing Unit (CPU) and the memory units are physically separated, inter-connected by a bus that works at lower frequencies than the CPU (the so-called Von Neumann architecture). This creates a bottleneck -usually known as "memory wall"- that enhances both the computing time and the computer energy consumption. Indeed, serious issues are envisaged in relation with the carbon footprint left by computers built under the Von Neumann paradigm and the massive use of ML algorithms [2]. Neuromorphic computing –which aims to mimic the architecture and the information processing mechanisms of the mammalian brain- offers energy saving for data intensive applications greater than any other computing paradigm [2, 3].

Memristors are defined as metal-insulator-metal structures capable to switch their electrical resistance between different levels -usually non-volatile- upon the application of external stimuli [3, 4]. For many memristive systems the resistance changes are quasi-analog and can, therefore, provide electrical adaptability, which is required to emulate the behavior of biological synapses [3]. This feature turns memristors into solid state devices with large potential to be one of the key building blocks for the development of neuromorphic hardware. Indeed, memristive devices have been used to fabricate physical NN -from simple perceptrons [5, 6] to recurrent [7] and convolutional [8] NN-, where the synaptic weights are linked to the conductivities of memristive devices that are organized in cross-bar arrays.

Memristive effects are usually found in oxides and linked to the electromigration of charged defects such as oxygen vacancies (OV), which could form conducting filaments or locally modulate the resistance of Schottky-like interfaces [4]. Other memristive mechanisms include charge trapping/detrapping effects [9, 10] or the modulation of tunnel barriers or Schottky interfaces by the direction of the polarization in the case of ferroelectric materials [11–15]. For some systems, memristive effects are concomitant with memcapacitive ones, that is the change of the device capacitance between different non-volatile states upon the application of electrical stimulation [16–27]. It was proposed that NN for character recognition formed by memcapacitors might significantly avoid energy dissipation by Joule effect and, therefore, they are expected to consume around $10^3$ less energy than those formed by memristors [28]. More recently, it was shown that reservoir computing systems for multimodal temporal information processing can be implemented by hardware using hafnium oxide-based memcapacitive networks, again with a several orders of magnitude reduction of the energy consumption in comparison with memristor arrays [29].

The systems displaying memcapacitance have been scarce and the few reported usually present relatively small ratios between high and low capacitive states ($C_{HIGH} / C_{LOW} \approx 10$ [16–24]), which might limit, for example, the functionality of circuity such as associate capacitive networks for character recognition [30]. An exception to this is the $La_{1/2}Sr_{1/2}Mn_{1/2}Co_{1/2}O_{3-x}$ (LSMCO) perovskite, which displays multi-mem properties (memristive and memcapacitive), with a large memcapacitance figure $C_{HIGH} / C_{LOW}$ of $\approx 100$ at 150 kHz [25].

LSCMO is a p-type perovskite that undergoes topotactic transitions and redox reactions [31], being possible to switch it between stable oxidized (x=0, more conducting) and reduced (x=0.62, more insulating) phases by the application of electrical stimulation [25, 27]. The corresponding topotactic redox reaction that describes the switching between both phases is $La_{0.5}Sr_{0.5}Mn_{0.5}Co_{0.5}O_3 \leftrightarrow La_{0.5}Sr_{0.5}Mn_{0.5}Co_{0.5}O_{2.38} + 0.31O_2(g)$.

When grown on $Nb:SrTiO_3$ (NSTO, n-type) single crystals, the NSTO/LSMCO interface forms a switchable n-p diode characterized, for oxidized LSMCO, by a thinner depletion layer with low resistance and high capacitance ($R_{LOW} / C_{HIGH}$) and, for reduced LSMCO, by a thicker depletion layer with high resistance and low capacitance ($R_{HIGH}/C_{LOW}$) [25]. Other topotactic redox systems with memristive behavior are $SrCoO_3$ [32–35] and $SrFeO_3$ [36–38]; however, none of these systems, to the best of our knowledge, display memcapacitive response.

Despite their multi-mem behavior with large memcapacitance, NSTO/LSMCO devices developed so far present two drawbacks [27]: i) the RESET transition (from $R_{LOW}/C_{HIGH}$ to $R_{HIGH}/C_{LOW}$ [39]) takes place at high voltages ($\approx +7$ V for (001) devices), which turns necessary to engineer the devices to reduce their operating voltages, ideally close to the 1 V threshold; ii) during the electroforming process, where the first LSMCO reduction takes place and oxygen is released to the ambient, structural damage and chemical changes occur, which electrically decouple the device active zone from the rest of the structure, hampering the possibility of integrating multiple devices in more complex architectures. Although it was shown that these structural and chemical changes can be reduced by switching the electrical stimulation from voltage to current (which self-limits the electrical power dissipation during the electroforming and RESET transitions) [27], the use of current stimulation is not the optimal solution to integrate these devices with standard -voltage

operated- circuitry, as this would require voltage-current conversion. Therefore, alternative strategies aiming at the operation of LSMCO-based devices with low voltages while maintaining the device integrity should be explored.

In the present work, we co-deposited LSMCO and $Sm_{0.3}Ce_{0.7}O_2$ (SCO, a well known excellent oxygen conductor [40]) in thin film form on NSTO single crystals, with (001) and (110) out-of-plane orientations, by Pulsed Laser Deposition (PLD) and using different growth conditions, in order to explore their multi-mem response. The two compositions have been deposited either as self-assembled composites [41, 42], where LSMCO/SCO are alternatively ordered at the nanoscale and SCO channels might favor an efficient, damage-free, oxygen exchange between the perovskite and the ambient, or in single phase films, where Ce(Sm) dopes the perovskite structure and could favour the topotactic redox transition at reduced voltages.

We obtained for (001)- and (110)-oriented NSTO/LSMCO-SCO/Pt devices significantly smaller RESET voltages compared to pristine LSMCO-based devices (≈ -64% and ≈ -45%, respectively), that maintain the multi-mem behavior with a large memcapacitive response up to $C_{HIGH}$ / $C_{LOW}$ ≈ 23. The (110) device showed, in addition, no substantial nanostructural damage or chemical modifications upon electrical cycling with voltage stimulation. We link these improvements to an increased LSMCO perovskite reducibility upon Ce(Sm) doping. Our work helps paving the way for the integration of voltage operated LSMCO-based devices in neuromorphic hardware.

## II. EXPERIMENTAL

LSCMO-SCO thin films were grown by Pulsed Laser Deposition from a ceramic mixed target (50/50 wt.%), using an excimer laser, on conducting $Nb:SrTiO_3$ (0.5 wt.%, NSTO) single crystals (Crystec) with out-of-plane (001) and (110) orientations. The deposition temperatures and background oxygen pressure were 800-840 ºC and 0.04 mbar, respectively. X-ray diffraction was performed with a Panalytical Empyrean diffractometer. Circular top Pt electrodes were deposited by sputtering and defined by optical UV lithography, with diameters between 45 $\mu$m and 500 $\mu$m. The memristive characterization was performed with a Keithley 2612 Source-Meter-Unit (SMU). The bottom electrode (NSTO) was grounded and contact was made on the top electrodes (Pt) with tungsten tips and a commercial probe-station. Memcapacitance was characterized with a LCR BK894 impedance analyzer, set to measure a parallel RC circuit. Capacitance-voltage loops were recorded at a frequency of 10 kHz. Impedance spectroscopy spectra were measured for frequencies in the range 100 Hz-500 kHz. High resolution Scanning Transmission Electron Microscopy with High Angular Annular Dark Field Detector (STEM-HAADF) measurements were performed with a FEI Titan G2 microscope with probe corrector (60-300 keV). In-situ chemical analysis was done by Energy Dispersive Spectroscopy (EDS, an Oxford instrument with AZtec software was used). TEM lamellas were prepared by Focused Ion Beam (FIB) in a Helios 650 Dual Beam Equipment.

### III. RESULTS

Figures 1(a) and (b) display x-ray diffraction (XRD) experiments performed on as-grown LSMCO/SCO thin films (LSMCO was in its oxidized phase [25]) on NSTO single crystals with (001) and (110) out-of-plane orientations, respectively. For both cases, XRD patterns are shown for growth temperatures between 800 and 880 °C and for laser repetition rates (f) between 0.08 and 1 Hz. The thickness of the oxide layers was in all cases ≈ 12 nm, as determined from x-ray reflectivity measurements (see Figure S1 of the Supplementary Information).

The Bragg-Brentano scans shown in the main panels of Figure 1(a), corresponding to LSMCO-SCO films grown on (001) NSTO, display the presence of (00l) perovskite reflections -indication of epitaxial growth-, while the (00l) SCO peaks develop only for laser repetition frequencies f ≤ 0.25 Hz. This suggests that the self-assembling of (001) LSMCO and SCO takes place at lower laser frequencies, while Ce and Sm are diluted in the perovskite structure otherwise. On the other hand, the Bragg-Brentano scans recorded for (110)-oriented samples, shown in Figure 1(b), display, in all cases, the sole presence of (hh0) perovskite diffractions peaks (the (220) reflection corresponding to SCO, not observed in the scans, is expected at $2\theta \approx 47.1°$), indicating that, for this orientation, Ce and Sm are prone to dilute in the perovskite structure, growing epitaxially on NSTO for the whole set of explored fabrication conditions.

In order to get structural and chemical information at the nanoscale, we have performed Transmission Electron Microscopy (TEM) on as-grown (001) and (110) oriented LSMCO/SCO samples. Figure 2 shows the case of an as-grown (001) sample deposited at 840 °C and 0.25 Hz (with self-assembled LSMCO and SCO phases according to the previously presented x-ray diffraction data).

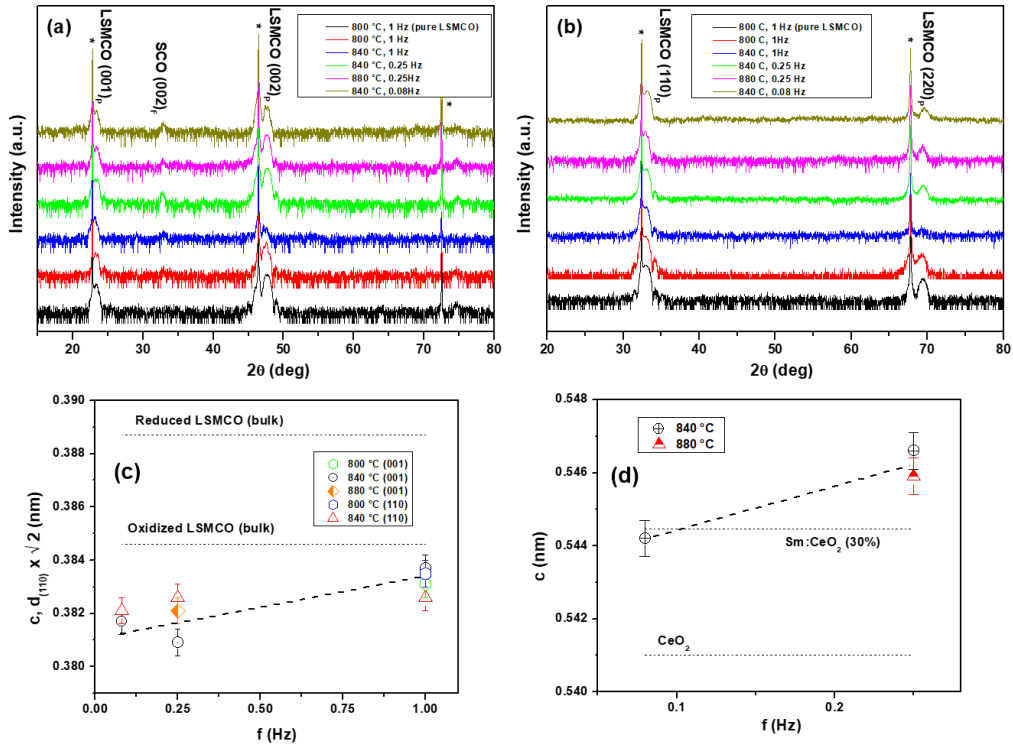

FIG. 1: (a), (b) Bragg-Brentano x-ray diffraction patterns corresponding to epitaxial LSMCO-SCO thin films grown on (001) and (110) NSTO, respectively, for different growth temperatures and laser repetition rates; (c) Evolution of the out-of-plane LSMCO lattice spacing for (001) (cell parameter c) and (110) LSMCO-SCO samples as a function of the laser frequency, for different growth temperatures. The pseudo-cubic bulk LSMCO cell parameter for both oxidized and reduced phases are shown as dotted lines; (d) Evolution of the out-of-plane SCO cell parameter for self-assembled (001) LSMCO-SCO samples as a function of the laser frequency, for two different growth temperatures. The bulk cell parameter of $CeO_2$ is shown with a dotted line.

Figure 2(a) displays a STEM-HAADF cross section, where the existence of SCO nanograins (with size ≈ 15-20 nm) located in a LSMCO matrix is evident. This is confirmed by the EDS line scan displayed in Figures 2(b) and (c), which also show that SCO grains are richer in Ce than the parent compound (nominally 70 at. %) and that Sm maintains a constant concentration both for LSMCO and SCO. The latter indicates that some Sm also dopes the LSMCO perovskite. The Fast Fourier Transforms (FFT) displayed in Figure 2(d) confirm that LSMCO and SCO phases are epitaxially coupled with epitaxial relations $(001)_P // (001)_F$ and $(110)_P // (100)_F$, where P and F refer to the LSMCO (pseudo-cubic) perovskite and SCO fluorite (cubic) structures, respectively.

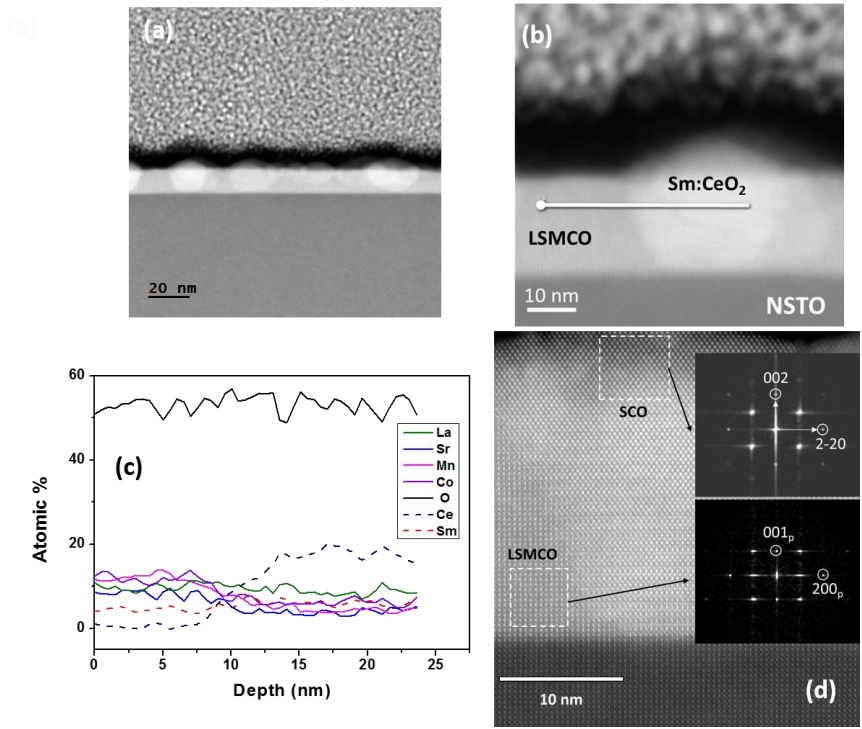

FIG. 2: (a) STEM-HAADF cross section corresponding to an as-grown self-assembled epitaxial (001) LSMCO-SCO film on NSTO. The presence of brighter SCO nanograins embedded in a LSMCO matrix is observed; (b) High resolution STEM-HAADF cross section of the same sample; (c) EDS line scans for La, Sr, Mn, Co, O, Ce, and Sm elements recorded on the same sample. The scan region is indicated with a white line in (b); (d) High-resolution STEM-HAADF cross section and Fast Fourier transforms (insets) performed on both LSMCO and SCO phases of the nanocomposite. The epitaxial coupling

Figure 3(a) shows a STEM-HAADF cross section corresponding to an as-grown (110) LSMCO-SCO film, grown at 840 ºC with a laser frequency of 0.25 Hz, displaying an homogeneous brightness indicating the absence of self-assembling between LSMCO and SCO and the dilution of Ce and Sm in the perovskite structure, in agreement with the x-ray diffraction scan shown before. This is confirmed by the EDS line-scan shown in Figures 3(b) and (c), which shows roughly constant concentrations for all the involved chemical species. However, a detailed inspection of the high resolution STEM-HAADF cross-section displayed in Figure 3(d), evidences some cationic inhomogeneity at the nanoscale (heavier atoms such as La, Sm and Ce are imaged as yellow/orange colours and lighter atoms such as Mn and Co as blue/violet colours).

Further structural information can be obtained from Figure 1(c), displaying the evolution of the out-of-plane LSMCO lattice spacing for (001) (c cell parameter) and (110) LSMCO-SCO samples, as extracted from the x-ray diffraction Bragg-Brentano scans of Figures 1(a) and (b), as a function of the laser repetition frequency. In all cases, the out-of-plane lattice spacing is lower than the values corresponding to bulk LSMCO (displayed as dashed lines), indicating that the perovskite unit cell is under tensile strain, as expected from the mismatch between NSTO (cubic structure with cell parameter a=0.3906 nm) and the oxidized LSMCO phase (with pseudo-cubic cell parameter a=0.3846 nm), in agreement with previous reports [43]. In addition, for (001) samples it is found that the out-of-plane lattice spacing increases with the laser frequency, reflecting the incorporation of Ce and Sm ions in the perovskite structure for the films grown at higher frequencies. We recall that rare earths such as Sm usually dope the A-site of the perovskite structure and that $Sm^{+3}$ ionic radii is smaller than the one of both $La^{+3}$ and $Sr^{+2}$ [44], indicating that unit cell changes are not driven by A-site doping with Sm (which might produce a cell contraction) but by B-site doping with Ce, where a cell expansion is expected given the higher ionic radii, for octahedral coordination, of Ce (in the range 0.97-1.14 ˚A, for +4 and +3 valences) with respect to Mn (in the range 0.64-0.96 ˚A, for +3 and +2 valences) and Co (0.61-0.90 ˚A, for +3 and +2 valences) [44]. The same reason explains, for laser frequencies of 0.08 and 0.25 Hz, the larger out-of-plane lattice spacing of (110) samples (Ce(Sm) are diluted) with respect to (001) samples (self-assembled). Figure 1(d) shows the evolution of the out-of-plane SCO cell parameter as a function of the laser repetition frequency, for (001) samples. It is seen that the out-of-plane cell parameter is in all cases larger than the one of the fluorite structure of CeO2, likely due either to the presence of some amount of (larger) Sm cations replacing Ce [44] or to the existence of compressive strain. These effects are more significant for films grown at higher laser frequencies.

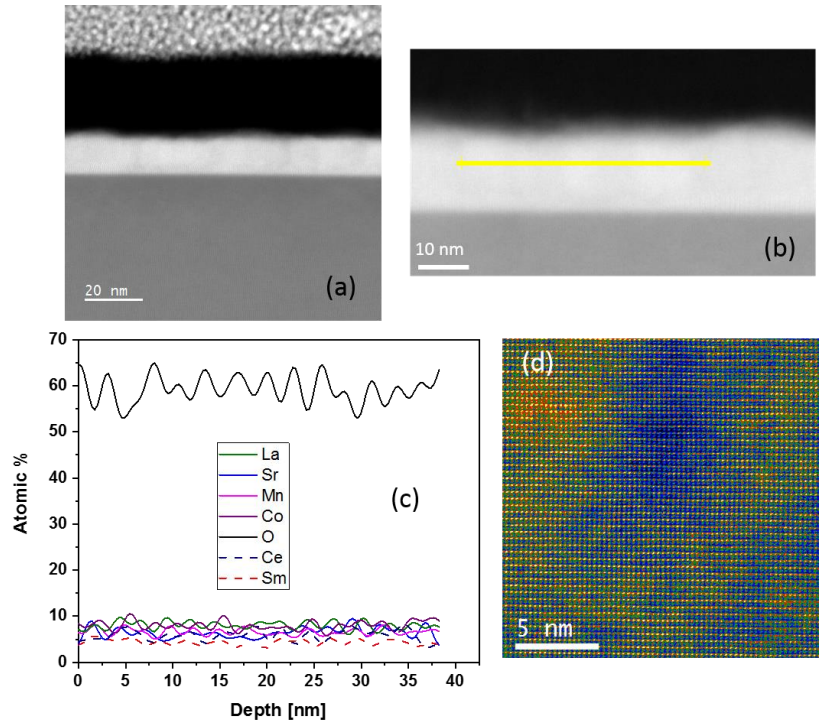

FIG. 3: (a) Low magnification STEM-HAADF cross section of an as-grown (110) LSMCO-SCO thin film on NSTO. The homogeneous brightness indicates the absence of self-assembly between LSMCO and SCO and the dilution of Ce and Sm in the perovskite structure; (b), (c) EDS line scans for La, Sr, Mn, Co, O, Ce, and Sm elements. The scan was performed in the zone indicated with a white line in (b); (d) High-resolution STEM-HAADF coloured cross section of the LSMCO-SCO thin film on (110) NSTO. Heavier atoms are imaged as yellow/orange colors, while lighter atoms are given blue/violet colors. The presence of local inhomogeneities is observed.

It is accepted that when two phases are co-deposited, their self-assembling depends on thermodynamic conditions such as surface, interface and elastic strain energies [41, 42]. For self-assembling, it is usually assumed that the two phases should present a significant difference in their interface energy with the substrate (usually defined as $\gamma_I$), in order to have different growth mechanisms. For instance, the phase with the lowest $\gamma_I$ should display a layer-by-layer growth mechanism (usually known as Stranski-Krastanov growth mode) while the one with the highest $\gamma_I$ must grow forming islands or columns (Volmer-Weber mechanism). For the case of out-of-equilibrium techniques such as PLD, we notice that kinetics could also control the self-assembling of different phases.

The observation of self-assembling of LSMCO and SCO for (001) films grown at lower laser frequencies can be explained either in terms of differences in their interface energy with the substrate (thermodynamic effect) or due to a low diffusivity of large Ce ions on the substrate surface (kinetic effect). The absence of self-assembling observed for higher growth rates - increased laser repetition frequency- indicates the presence of smaller sized islands due to shorter diffusion time of Ce(Sm) adatoms in the substrate surface in between laser pulses. For (110) LSMCO-SCO structures, the absence of self-assembling between both phases, for all the explored growth conditions, might indicate either that NSTO/LSMCO and NSTO/SCO interface energies are closer or that Ce(Sm) diffusivity is increased for this orientation, leading to phase intermixing and the dilution of Ce(Sm) in the perovskite LSMCO structure.

We focus now on the memristive response of (001) and (110) NSTO/LSMCO-SCO/Pt microdevices. In all cases, NSTO was grounded and voltage stimulation was applied to the top Pt electrode. The area of the devices was $22.3 \times 10^3\ \mu m^2$. The virgin resistance states were 450 kΩ and 340 kΩ for (001) and (110) devices, respectively, which dropped to 200 Ω and 120 Ω, respectively, after the application of electroforming -6 V pulses (1 ms wide). The inset of Figure 4 (a) shows the electroforming process for a (001) device.

After forming the devices, we recorded simultaneously dynamic current-voltage curves (I-V) and remnant resistance loops (or Hysteresis Switching Loops, HSL). For the former, a pulsed ramp of write voltage pulses (also 1 ms wide) were applied and the current was measured during the application of the pulses. In between write pulses, we measured the remnant resistance with a small 100 mV bias, in order to construct the HSL [45].

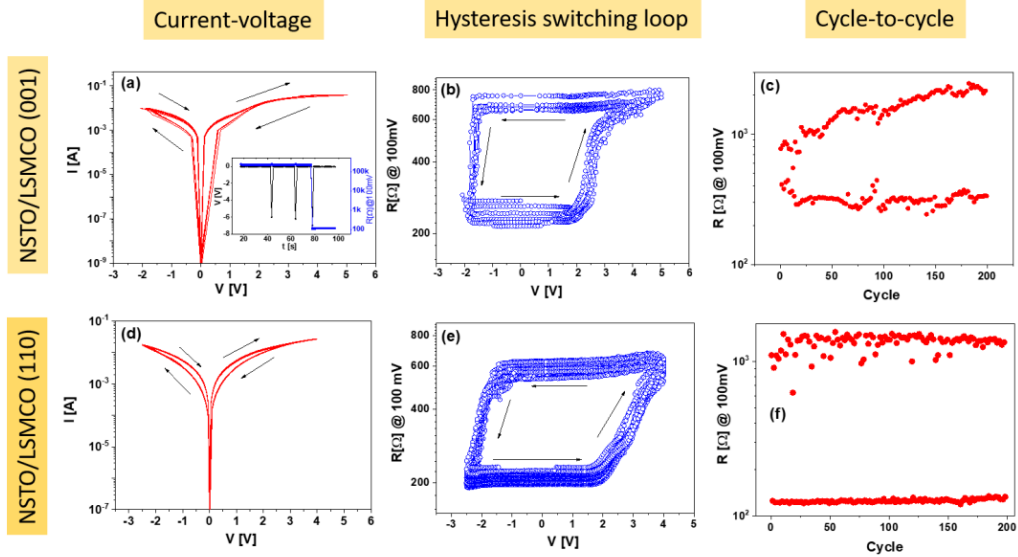

FIG. 4: (a), (d) Dynamic I-V curves recorded for (001) and (110) NSTO/LSMCO-SCO/Pt devices, respectively. The (001) device displayed initially LSMCO and SCO self-assembled phases. The inset in (a) shows the electroforming process, achieved after the application of a train of pulses up to -7 V; (b), (e) Hysteresis Switching Loops corresponding to the same devices, recorded simultaneously with the I-V curves shown in (a) and (d); (c) and (f) Cycle-to-cycle variability tests performed on (001) and (110) NSTO/LSMCO-SCO/Pt devices, respectively, for 200 cycles.

Figures 4 (a) and (d) show the dynamic I-V curves for (001) and (110) NSTO/LSMCOSCO/Pt devices, respectively, while Figures 4 (b) and (e) show the corresponding HSLs. In the case of the (001) device, LSMCO and SCO were initially self-assembled. The transition from $R_{HIGH}$ to $R_{LOW}$ (SET process) is achieved after the application of ≈ -1.5 V and ≈ -1.8 V pulses, for (001) and (110) devices, while the inverse transition (RESET process) is obtained upon the application of ≈ +2.5 V and ≈ +3 V pulses, respectively. The RESET voltages are significantly smaller in comparison with the ones measured on (001) and (110) NSTO/pure LSMCO/Pt devices ( ≈ +7 V and ≈ +5.5 V respectively, see the Supplementary Information, Figure S2). Also the SET voltages are smaller than the values observed on (001) and (110) NSTO/pure LSMCO/Pt devices (≈ -1.8 V and ≈ -2.5 V, respectively, see Figure S2). The ON-OFF window -defined as the ratio $R_{HIGH}/R_{LOW}$ was ≈ 3 for both (001) and (110) NSTO/LSMCO-SCO/Pt devices, respectively, somewhat smaller values than those found for pure LSMCO-based devices (≈ 8 and ≈ 7 for (001) and (110) orientations, respectively, see Figure S2). Figures 4 (c) and (f) display cycle-to-cycle (ctc) variability tests for NSTO/LSMCO-SCO/Pt (001) and (110) devices, respectively. To obtain these curves, single SET and RESET pulses were applied for 200 cycles. It is seen that for the (001) device the $R_{HIGH}$ state shows a drift towards higher values upon cycling while $R_{LOW}$ remains stable. For the (110) device both $R_{HIGH}$ and $R_{LOW}$ are stable upon cycling. This is confirmed by calculating the cumulative probabilities linked to both states, as displayed on Figure S3, which show steeper $R_{HIGH}$ and $R_{LOW}$ curves for the (110) device, indicating a lower cycle-to-cycle variability for this orientation. Retention times up to $10^3$ s are displayed in Figure S4.

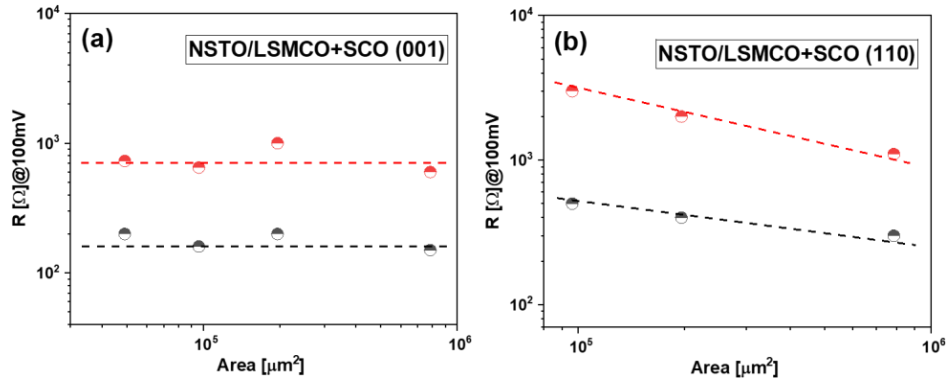

FIG. 5: (a), (b) Evolution of $R_{LOW}$ and $R_{HIGH}$ as a function of the device area for (001) and (110) NSTO/LSMCO-SCO/Pt devices, respectively.

We address now the possible origin of the difference between the more stable memristive response of (110)-oriented devices against the electrical behavior of (001) ones. First, we compare in Figure 5 the evolution of both $R_{HIGH}$ and $R_{LOW}$ with the device area, for (001) (panel (a)) and (110) (panel (b)) orientations. It is seen that in the latter case both $R_{HIGH}$ and $R_{LOW}$ states increase as the area is reduced, while in the former case the $R_{HIGH}$ and $R_{LOW}$ states are both area-independent. This indicates that in the case of (110) devices the current is distributed along their complete area during electrical cycling, with the device remaining undamaged by the electrical stimulation. This is confirmed by the optical top-view image taken after the application of electrical stress, shown in Figure S5, which displays no observable damage. For the (100) case, instead, the constant values of $R_{HIGH}$ and $R_{LOW}$ for different device areas indicate the existence of electrical decoupling upon forming between the zone where the electrical probe lands and the rest of the device, as we previously reported for pure LSMCO devices [25, 27]. The damage is related to the release of oxygen to the ambient during electroforming and LSMCO reduction, accelerated by self-heating effects which also fuse the Pt top electrode in contact with the electrical tip (a zone with a diameter of ≈ 15 $\mu$m is observed from the optical top view image displayed in Figure S5). The same figure shows a ring around the electrical tip landing zone, where both LSMCO and Pt are expelled [25]. This process does not take place for (110) devices as LSMCO reduction is favoured trough oxygen migration along the [001] direction, which are easy migration channels bridging both electrodes, as shown for oxide thin films displaying reduced brownmillerite-like structure such as $SrCoO_3$ [32–35], $SrFeO_3$ [36–38] and LSMCO [25, 27].

Further info is obtained from the analysis of STEM-HAADF images taken on stressed NSTO/LSMCO-SCO/Pt (001) (initially self-assembled) and (110) devices, as shown in Figures 6 and S6, respectively. Figure 6(a) shows a STEM-HAADF cross-section corresponding to a (110) device, where the TEM lamella was prepared below the landing zone of the electrical tip. The homogeneous brightness evidences the dilution of Ce and Sm in the perovskite structure. This is confirmed in Figure 6(b), which displays EDS line scans, recorded across the region indicated in panel (a), showing the co-existence of LSMCO cations with Ce(Sm), with roughly constant concentrations along the (100) direction (parallel to the substrate interface). Figures 6(c) and (d) repeat the previous analysis in a lamella prepared outside the landing zone of the electrical tip. In this case, LSMCO and SCO remain as separated phases, as infered both from the presence of some STEM-HAADF bright-dark contrast (panel (c)) and the EDS line scans, which clearly show the presence of an SCO grain at the center of the scan (panel (d)). This indicates that, as expected, Ce and Sm dilution upon electrical cycling occurs only in the electrically active zone of the device. The STEM-HAADF and EDS analysis of a stressed (110) device is shown in Figure S6, and displays no significant differences with respect a virgin device; that is, the LSMCO phase remains doped with Ce(Sm).

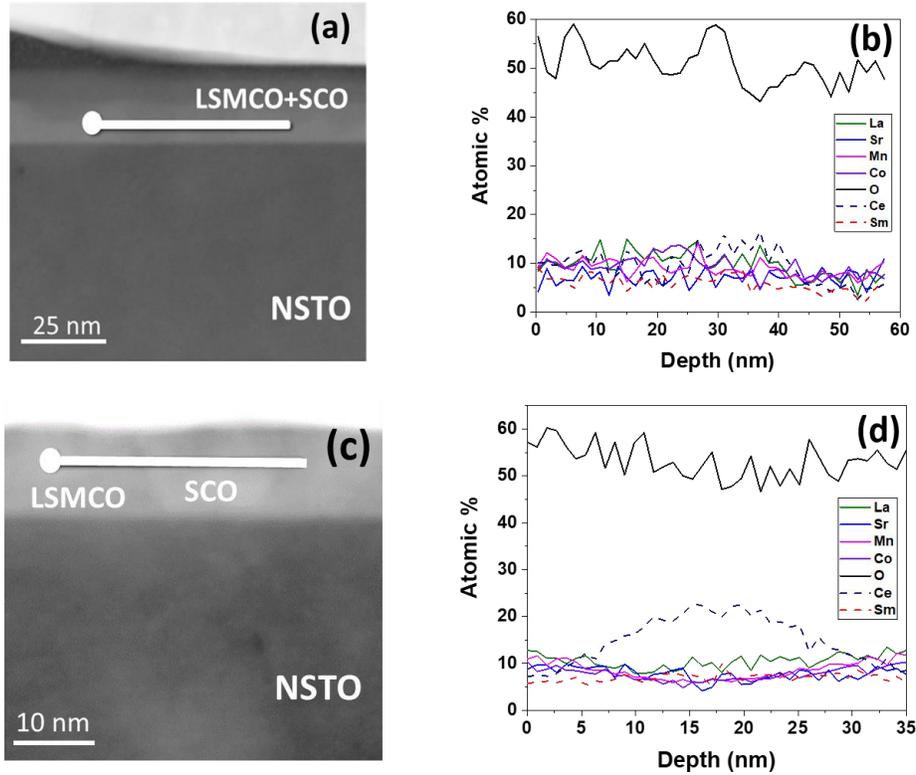

FIG. 6: STEM-HAADF cross sections corresponding to an initially self-assembled (001) NSTO/LSMCO-SCO/Pt device after electrical stress. The lamellas were prepared in the active area below the contact tip (a) and outside the active area (c); (b) and (d) EDS line scans for La, Sr, Mn, Co, Ce, and Sm elements. The oxygen quantifications were performed by EELS. The scans, indicated by white lines in (a) and (c), indicate the dilution of Ce(Sm) in the perovskite structure in the active area of the device.

Based on the previous analysis, it can be concluded that the instability of the memristive response of the initially self-assembled (001) device is related to the progressive mixing of LSMCO and SCO phases upon electrical cycling. The active zones of both (001) and (110) devices end up being LSMCO doped with Ce(Sm), which explains the closeness in their SET and RESET voltages, with lower values than those found for pure LSMCO.

We turn now to the memcapacitive response of the devices. Figure 7 shows a remnant capacitance vs. writing voltage, measured with a LCR meter in a (110) NSTO/LSMCO-SCO/Pt device. The presence of hysteresis is evident, with a $C_{HIGH}$ ($C_{LOW}$) state, concomitant with the $R_{LOW}$ ($R_{HIGH}$) state, of ≈ 23 nF (≈ 1 nF). The transition from $C_{HIGH}$ to $C_{LOW}$ is obtained after the application of +3.5 V, in good agreement with the resistive SET voltage extracted from Figure 4, while the opposite transition (resistive RESET) is observed for -2.3 V, again in good agreement with Figure 4. The slight differences in the transition voltages between memristive and memcapacitive loops are a consequence of the use of different write pulse time-widths, due to instrumental limitations. The $C_{HIGH}/C_{LOW}$ ratio is ≈ 23, larger than the figures reported so far for other memcapacitive systems ($C_{HIGH}/C_{LOW}$ ≤ 10 [16–25, 27]). We also notice that the capacitance change is analog-like, with ≈ 20 intermediate states stabilized between $C_{HIGH}$ and $C_{LOW}$ states. This underscores the potential of variable capacitance states for implementing synaptic weights in physical neural networks developed from memcapacitor cross-bar arrays.

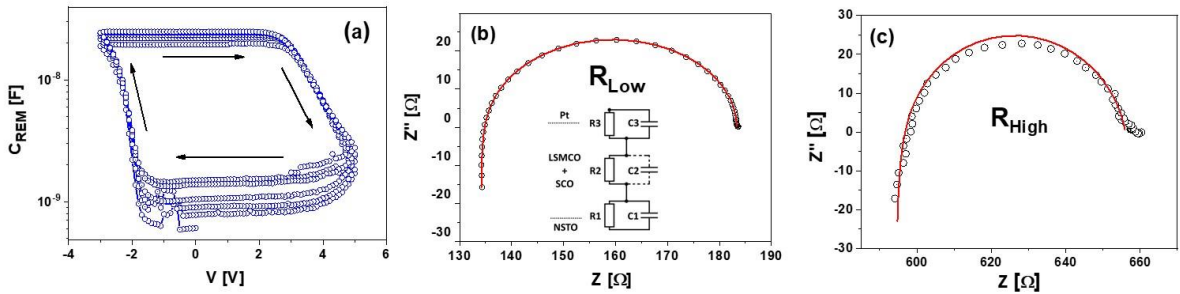

FIG. 7: (a) Remnant capacitance vs. writing voltage corresponding to a (110) NSTO/LSMCO-SCO/Pt device. Two clear capacitive states are observed, with $C_{HIGH}$ corresponding to $R_{LOW}$ and $C_{LOW}$ corresponding to $R_{HIGH}$; (b), (c) Impedance spectra measured for $R_{LOW}$/$C_{HIGH}$ and $R_{HIGH}$/$C_{LOW}$ states, respectively. The inset in (b) shows the equivalent circuit used to fit the spectra. The fittings are shown in red lines.

The memcapacitive effect was also inferred from impedance spectroscopy experiments performed on both $R_{LOW}$ and $R_{HIGH}$ states, as shown in the Nyquist plots of Figures 7(b) and (c). The spectra were fitted by assuming the equivalent circuit displayed in the inset of Figure 7(b). The elements $R_1$ and $C_1$ correspond to the NSTO/LSMCO-SCO interface, $R_3$ and $C_3$ to the LSMCO-SCO/Pt top interface and $R_2$ and $C_2$ to the LSMCO-SCO layer in between both interfaces [25, 27]. The branch containing $C_2$ is neglected as it presents a much higher impedance respect to $R_2$ [27]. The fitted values for the circuit elements are displayed in Table S1. The most remarkable feature is that the memcapacitive effect is dominated by the $C_3$ element, which was attributed to the (variable) capacitance of the n-p diode formed at the NSTO/LSMCO interface, which changes between 41 nF for oxidized LSMCO and 3 pF for reduced LSMCO. The memcapacitive response of the (001) NSTO/LSMCO-SCO/Pt device is shown, for completeness, in Figure S7. The $C_{HIGH}/C_{LOW}$ ratio was ≈ 13 and we found that both $C_{HIGH}$ and $C_{LOW}$ are around two orders of magnitude lower than in the (110) devices. We relate this to the different effective areas of the devices, comprising the complete top electrode in the case of (110) device and only the (decoupled) landing zone of the tip, around 100 times lower, in the case of the (001) device (recall Figure 5 and the corresponding discussion).

IV.    DISCUSSION AND CONCLUSSION

The results presented in the previous section show that LSMCO-based devices modified with SCO maintain the multi-mem behavior of the pristine material but can be operated at lower voltages; in particular we obtained a significant reduction of the RESET voltage of ≈ -64 % ((001) devices) and -45 % ((110) devices) with respect to the pristine perovskite. In addition, a large memcapacitance with ON/OFF ratios up to 23 is demonstrated.

We disclosed some relevant differences between (001) and (110) devices, which we address now. In the first place, our analysis shows that (001) devices -where LSMCO and SCO are self-assembled during growth (see the sketch of Figure 8(a))- suffer some structural damage upon forming, that electrically decouples the active zone of the device from the rest of the structure, while (110) devices -where Ce(Sm) is diluted in the perovskite structure, see the sketch of Figure 8(b)- maintain their integrity. The presence of damage in (001) devices is related to oxygen release to the ambient during the perovskite reduction and indicates that the migration of oxygen out of the device is not performed optimally trough SCO channels, as expected and reported for other systems such as SCO-SrTi$O_3$ composites [40]. A possible explanation of this is given by Dou et al. [46], where they analyze the oxygen migration process in CeO$_2$ nanostructures with different morphologies. Is it found that a "columnar scaffold" is the optimal morphology that maximizes oxygen movement in the out-of-plane direction. Instead, an island-like morphology, similar to our SCO nanograins, is reported as detrimental to the oxygen transport in the direction perpendicular to the substrate surface, necessary in our case for LSMCO reduction and the concomitant $O_2$ release.

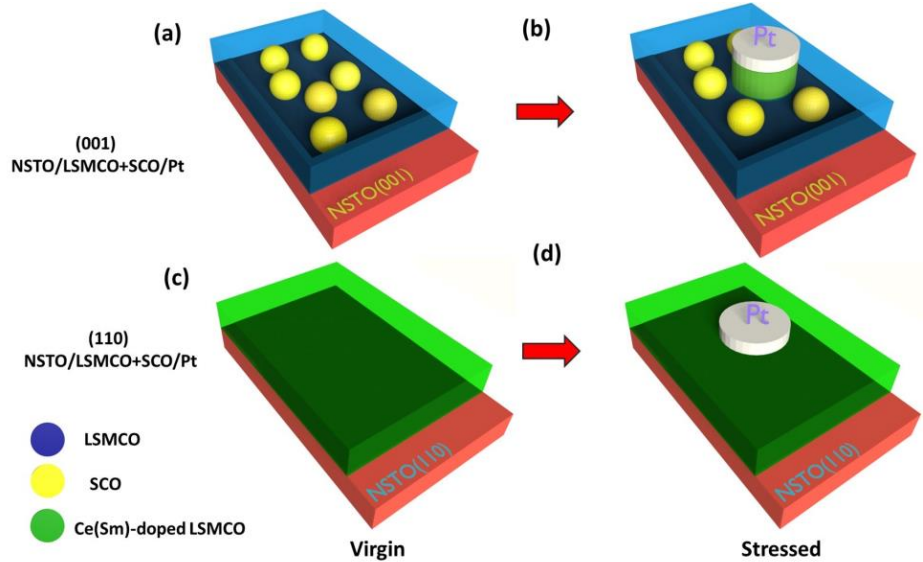

FIG. 8: Sketch depicting the nanostructures of as-grown (001) (a) and (110) (c) NSTO/LSMCO-SCO films. In the first case both LSMCO and SCO phases are self-assembled, while in the second case Ce(Sm) dope the LSMCO perovskite. Panel (b) displays the (001) device after the application of the electrical stress, where, in the zone below the top Pt electrode, LSMCO and SCO intermix and Ce(Sm) dope the perovskite. In the case of the electrically stressed (110) device (panel (d)), no significant changes are seen in relation to the virgin device.

We also observed, for the (001) device, that the memristive response was somewhat unstable, with a drift of the $R_{HIGH}$ state upon cycling. In addition, the STEM-HAADF-EDS analysis performed on an electrically stressed (001) device shows the dilution of Ce(Sm) in the perovskite LSMCO structure (see the sketch of Figure 8(c)). We relate the electrical instability to the progressive mixing of LSMCO and SCO phases upon electrical cycling. Once the dilution takes place, both (001) and (110) devices present Ce(Sm)-doped LSMCO and display a memristive behavior with similar figures (namely SET and RESET voltages and $R_{HIGH}/R_{LOW}$ ratios). In particular, the fact that the RESET voltage is significant smaller than the case of pristine LSMCO indicates that the incorporation of Ce(Sm) enhances the reducibility of LSMCO. This is in agreement with previous reports on other other perovskites such as LaCo$O_3$ [47] and LaMn$O_3$ [48] where it is shown that doping with Ce improves their reducibility and enhances the catalytic performance for benzyl alcohol and CO oxidation, respectively. For the case of LSMCO thin films, the enhanced reducibility of Ce-doped samples can be related to a steric effect related to the unit cell expansion produced by Ce incorporation to the perovskite structure, as previously shown by van den Bosch et al. in LSMCO thin films grown with different strains and unit cell deformations [43]. Additionally, strain and Sr doping were also shown to tune the threshold voltage related to the perovskite-brownmillerite transition in La$_{1-x}$Sr$_x$CoO$_3$ thin films grown on different substrates [49]. This suggests a route to further reduce the operating voltages of LSMCO-based devices by fine controlling unit cell deformations either by strain engineering and/or the use of alternative dopants.

To conclude, we have disclosed a simple way of lowering the operating voltages of (110) LSMCO films by modifying then with Ce(Sm), while maintaining the multi-mem behavior and the structural and chemical integrity of the devices. This strategy surpasses the previously reported use of current stimulation in pure LSMCO devices [27] as it allows the use of voltage stimulation, facilitating the integration with other electronic components without the need of current-voltage conversion. The results reported here could help paving the way for the implementation of memristive and memcapacitive LSMCO in neuromorphic electronics.

**SUPPORTING INFORMATION AVAILABLE**

Additional x-ray reflectivity experiments performed on NSTO/LSMCO-SCO thin films with (001) and (110) out-of-plane orientation; electrical characterization of a pure LSMCO sample on NSTO, with top Pt electrode; calculated cumulative probabilities of the cycleto-cycle variability experiments shown in Figure 4; retention times corresponding to the same NSTO/LSMCO-SCO/Pt samples; optical top-view images of NSTO/LSMCO-SCO/Pt devices; STEM-EDX data corresponding to a stressed NSTO/LSMCO-SCO/Pt (110) device; impedance spectroscopy and memcapacitive effect recorded on a NSTO/LSMCO-SCO/Pt (001) device;

tables showing the values of the fitted parameters of impedance spectroscopy spectra of NSTO/LSMCO-SCO/Pt samples with (001) and (110) orientations.

**ACKNOWLEGMENTS**

We acknowledge support from ANPCyT (PICT2019-02781 and PICT2020A-00415) and EU-H2020-RISE project MELON (Grant No. 872631). This publication is also a result of the project "Memcapacitive elements for cognitive devices", project number 040.11.735, which was financed by the Dutch Research Council (NWO). We thank U. Lu¨ders and J. Lecourt for the preparation of the PLD targets.